# High temperature *in situ* SEM assessment followed by *ex situ* AFM and EBSD investigation of the nucleation and early growth stages of Fe-Al intermetallics

T. Sapanathan[1,*], I. Sabirov[2], P. Xia[2,3], M.A. Monclús[2], J.M. Molina-Aldareguía[2], P.J. Jacques[1], A. Simar[1]

[1] UCLouvain, Institute of Mechanics, Materials and Civil Engineering, IMAP, B-1348 Louvain-la-Neuve, Belgium

[2] IMDEA Materials Institute, Calle Eric Kandel 2, 28906 Getafe, Madrid, Spain

[3] Universidad Politécnica de Madrid, E.T.S. de Ingenieros de Caminos, 28040 Madrid, Spain

* Corresponding author e-mail: thaneshan.sapanathan@uclouvain.be

Abstract

A dedicated *in situ* heating setup in a scanning electron microscope (SEM) followed by an *ex situ* atomic force microscopy (AFM) and electron backscatter diffraction (EBSD) is used to characterize the nucleation and early growth stages of Fe-Al intermetallics (IMs) at 596 °C. A location tracking is used to interpret further characterization. *Ex situ* AFM observations reveal a slight shrinkage and out of plane protrusion of the IM at the onset of IM nucleation followed by directional growth. The formed interfacial IM compounds were identified by *ex situ EBSD*. It is now clearly demonstrated that the θ-phase nucleates first prior to the diffusion-controlled growth of the η-phase. The θ-phase prevails the intermetallic layer.

Different Fe-Al intermetallics (IMs), also known as iron aluminides, can form by reaction implying diffusion in the Fe-Al binary system. The exact reaction scheme and the observed coexistence of several IMs when considering the reactions occurring at an interface between Fe and Al still need to be unveiled. Specific attention has been paid in literature to $Fe_3Al$ - β1 (< 50 at.% Al), which offers good resistance to sulfidation, oxidation at high temperature and mechanical wear, while presenting a relatively high ductility among all the Fe/Al IMs and a lower density than steel [1]. As summarized in Table 1, the other Al-rich (≥ 50 at.% Al) Fe-Al IMs are FeAl - β2, $FeAl_2$ - ζ, $Fe_2Al_5$ -η, $Fe_4Al_{13}$ – θ (also referred to as $FeAl_3$ in some early literature) and $Fe_5Al_8$ - ε (only stable at temperatures above 1095 °C [2]). The β2 and ζ phases form via decomposition of the ε-phase resulting in a lamellar microstructure [3-5]. β2 and ζ have thus also gained some attention for providing improved mechanical properties similar to the more commonly known TiAl-based alloys [5]. The other two Al-rich IMs, η- and θ- phases, have very low reported fracture toughness of 0.82 MPa·m$^{1/2}$ and 0.97 MPa·m$^{1/2}$, respectively [6]. Such low toughness IMs are undesirable for many engineering applications.

It is well established that the brittleness of the IM phases increases with increasing at.% Al content. However, the Al-rich IMs (especially the η- and θ- phases) spontaneously form during various manufacturing processes, e.g. welding, coating, sintering, aluminum casting with ferrous inserts, hot-dip aluminizing, etc. [7-10]. Therefore, understanding the formation mechanism of the η- and θ- phases is crucial to avoid their undesirable formation during manufacturing processes. During these processes, the local temperature at the interface rapidly changes with time, which could play a significant role on the sequence of IM phase nucleation.

Table 1: Lattice structure, stability range, standard enthalpy and entropy change at 298 K and Gibb's free energy changes at 973 K for IMs of the Fe-Al binary system [5, 11-13]*

| Intermetallic phases | Lattice structure | Stoichiometry of Al (at.%) | $\Delta H_{298}$ (J mol$^{-1}$) | $\Delta S_{298}$ (K$^{-1}$ mol$^{-1}$) | $\Delta G_{973}$ (J mol$^{-1}$) |
|---|---|---|---|---|---|
| $Fe_3Al$ (β1) | DO3 | 23 - 34 | -57372 | 28 | -4827 |
| FeAl (β2) | BCC (order B2) | 23 - 55 | -51240 | 51 | -11090 |
| $FeAl_2$ (ζ) | Triclinic | 66 - 66.9 | -81900 | 73.3 | -16999 |
| $Fe_2Al_5$ (η) | Orthorhombic | 70 - 73 | -194040 | 166.7 | -19636 |
| $Fe_4Al_{13}$ (θ) | Monoclinic | 74.5 - 76.5 | -112560 | 95.6 | -22869 |

* The stable phases at ambient temperature are listed in ascending at.% of Al

To date, many *ex situ* analyses have been performed to understand the formation of Fe-Al IMs using dedicated set-ups and post treatments [9, 13-16]. In most of these studies, solid iron or steel was dipped into liquid or semi-solid Al at different temperatures and for different times. Early work [9] mentioned that the η-phase is dominating, and its growth follows a parabolic law for temperatures ranging from 715 °C to 944 °C. In another study [17], the Fe-Al intermetallic growth kinetics parameters were determined for temperatures ranging from 500 °C to 1100 °C based on hot dipping experiments. Al diffusion was considered as the main mechanism during the η-phase growth, while Fe diffusion was also suggested as an active mechanism based on the emergence of Kirkendall voids in the $Fe_2Al_5$ layer [17]. It should be noted that in such *ex situ* analyses, the onset of nucleation and the growth of various IMs could not be captured due to their extremely fast kinetics which requires a sophisticated experimental setup. Therefore, the early stages of IM nucleation and growth in the Fe-Al system remain mostly unexplored.

Recently, *in situ* high temperature experiments have become an important tool to elucidate the behaviour at moving interfaces. Austenite - ferrite moving interface during high temperature treatment was successfully characterized for steel using *in situ* EBSD analysis [18, 19]. *In situ* X-ray absorption tomography was used to characterize the IM growth at 700 °C (i.e. after Al melting), allowing the observation of tongue-like growth of the η-phase [20]. Synchrotron X-ray radiography was also used to characterize the IM growth at the liquid Al/ solid Fe interface [21]. The authors reported the formation of the η-phase at 850 °C, whereas a dotted θ-phase and a needle-like θ-phase formed during solidification [21]. The nature of the phases was confirmed by electron back scatter diffraction (EBSD) analysis of the heat-treated samples after significant growth. The early sequence of formation of the phases could thus not be accurately captured during this *in situ* X-ray radiography study as this technique did not allow to distinguish between the various Fe-Al phases nor track phases location. Owing to *in situ* X-ray diffraction (XRD) measurements, the formation of η- and ζ- phases below the melting temperature of aluminum during a slow heating at a rate of 10 C°/min was also emphasized [22]. Moreover, *in situ* thermal treatment in a transmission electron microscope (TEM) identified that a needle like θ-phase formed during soaking at 350 °C [23]. The TEM observations suggested that the θ-phase forms first during IM formation at the Fe-Al interface, while other studies are still not conclusive [20-22]. This scenario may be due to the temporal/spatial resolutions of the *in situ* devices used in those experiments, as well as their thermal inertia.

The main objective of the present work is to shed more light on the early stages of IMs nucleation and growth in the Fe-Al system. A controlled *in situ* heat treatment within a SEM chamber was followed by detailed microstructural characterization of the IM sites. Roll bonding (RB) was used to prepare bonded Al-Fe joints following the procedure described in [24]. DP 600 steel and AA6061-T6 aluminum having the dimensions of 200 x 80 x 0.9 mm$^3$ (L × W × H) and 200 x 80 x 3 mm$^3$, respectively, were stacked in the order of DP600 / AA6061-T6 / DP600 and wrapped in a steel foil to avoid relative sliding during the rolling process. The chemical compositions of the DP600 and AA6061-T6 are provided in Table S1 in Supplementary. The sample was subjected to four rolling passes to obtain a thickness reduction ratio of 50%. 200 μm-thick specimens were then extracted from the transverse cross section of the Roll-bonded (RB) sample for the *in situ* heating experiment. Such thin samples were used for the *in situ* SEM experiments to eliminate the effect of thermal gradients along the thickness of the sample during heating. The prepared interface is shown in Fig. 1a revealing a clean interface without any visible defects or any intermetallic compounds at the interface. In the selected zone, two micropores observed on the surface were used as position tracking markers during the *in situ* experiment (Fig. 1a) and for subsequent *ex situ* AFM and *ex situ* EBSD analyses.

To identify the activation temperature required for the nucleation and growth of IMs, differential scanning calorimetry (DSC) experiments were performed with a PERSEUS® STA 449 F1/F3 equipment. The 2 × 2 × 2 mm$^3$ samples obtained from the RB specimen were used in this experiment. An identified peak corresponding to the intermetallics formation is normalized by the formed total intermetallic volume (0.00138 mm$^3$) and given in Fig. 1c. It is worth noting that, due to the presence of multiple IMs, the DSC results are not normalized to the unit mass. The DSC experiment showed that the temperature required to start forming the IM is about 577 °C. Initial trials of the *in situ* SEM experiments were thus carried out at different temperatures ranging from 570 °C to 600 °C with 10 °C increments (and a soaking time of 60 sec). From the *in situ* SEM experiments, it was identified that a temperature of 596 °C is required for the nucleation and growth of IM compounds, which can be recorded with a reasonable resolution (see the Supplementary video). The temperature difference from the DSC scan and the *in situ* SEM observation can be attributed to the location of the thermocouple. Indeed, the DSC sample was placed inside an $Al_2O_3$ crucible in a furnace made of SiC chamber, and the thermocouple was located below that crucible; whereas in the SEM *in situ* experiment, the thermocouple was touching the sample surface (see Fig. 1b).

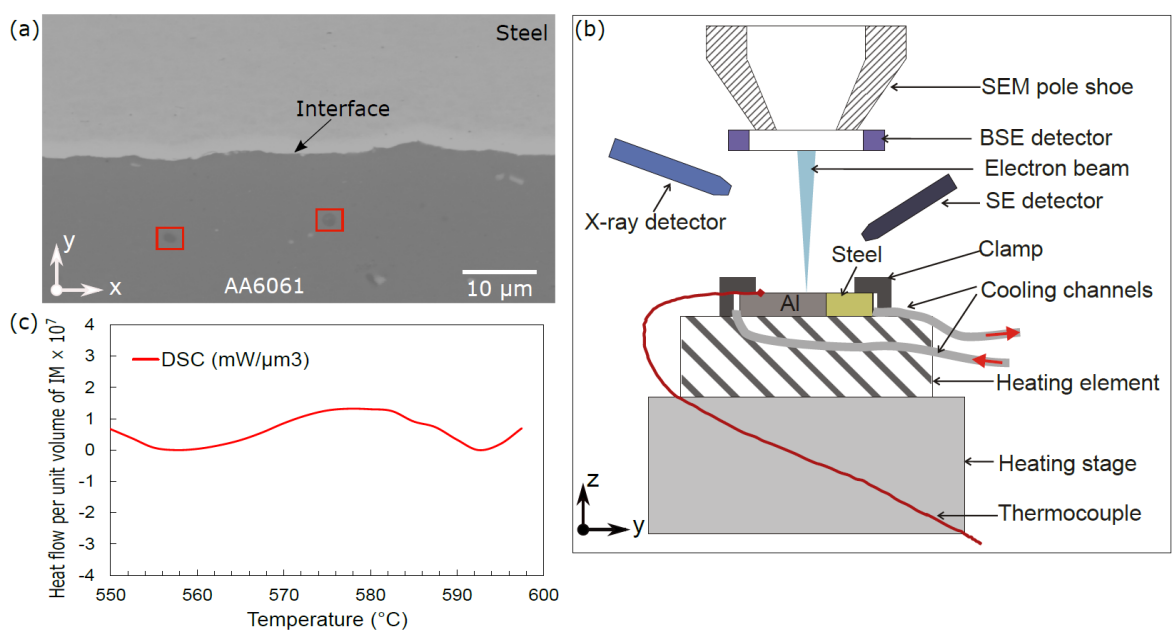

Fig. 1. (a) SEM micrograph of the interface of one RB sample used for the *in situ* experiments. The red squares indicate two micropores in the Al alloy that were used for position tracking during the *in situ* experiments and subsequent *ex situ* AFM and EBSD analysis. (b) Schematic illustration of the setup used for the high temperature *in situ* experiments inside the SEM chamber. (c) Heat flow behavior specific to the unit volume of intermetallic obtained from DSC measurement.

A ZEISS SEM EVO MA15 equipped with a heating stage, which is part of the Hysitron PI88 picoindenter, was used to carry out the *in situ* thermal treatment (Fig. 1b). Temperature was actively measured by a thermocouple placed in contact with the sample surface and feedback-controlled via a Lakeshore 336 PID temperature controller to ensure maximum thermal stability at the desired values (within ± 0.01 ˚C). The temperature with time plot is given in the supplementary (Fig. S1). The microscope was operated at 18 kV. The electron beam scanned over the 69.2 µm × 51.9 µm surface area perpendicular to the interface (y direction), and secondary electron (SE) signal was continuously recorded (at 12 frames/min) during heating. The pixel size of the SEM images was 67.4 nm. This allowed to obtain SEM images of sufficient quality while keeping track of microstructure evolution during heating. After the heating stage, the topography of the nucleated IMs at the interface was characterized using an XE-150 Park´s AFM system fitted with Acta® cantilever scanning tips. The AFM was operated in non-contact mode at a scan rate of 0.5 Hz to obtain topological images with typical scan areas of 10 × 10 µm$^2$ at 512x512 pixels resolution. EBSD analysis of the same region (after fine polishing) was also performed using FEI Quanta TM Helios NanoLab 600i, equipped with a NordlysNano detector controlled by the AZtec Oxford Instruments Nanoanalysis software. The EBSD data were acquired at an accelerating voltage of 18 kV, a working distance of 8 mm, a tilt angle of 70° and a step size of 100 nm in a square scan grid mode.

Figures 2a-d illustrate the sequence of IM nucleation and growth during the *in situ* heating experiments. Image analysis was performed using the Image J software to quantitatively measure the extent of the IM growth at the interface (Fig. 2e). As shown in Fig. 2e, the IM nucleates at various locations during the first 10 s of soaking at 596 °C, and a fast growth occurs during the first 40 s of soaking (for Locations B to F marked in Fig. 2d), followed by saturation of the growth. After the onset of nucleation (at t=0), the initial IM growth rate is ~0.183 µm/s. However, when the growth is delayed by about 10 s, the corresponding total IM growth rate at the beginning slowed down to ~0.081 µm/s (indicated by the curve corresponding to Location A in Fig. 2e). It is worth noting that in Location A, the intermetallic still continues its growth at the end of the experiment (60 s), although at a lower rate.

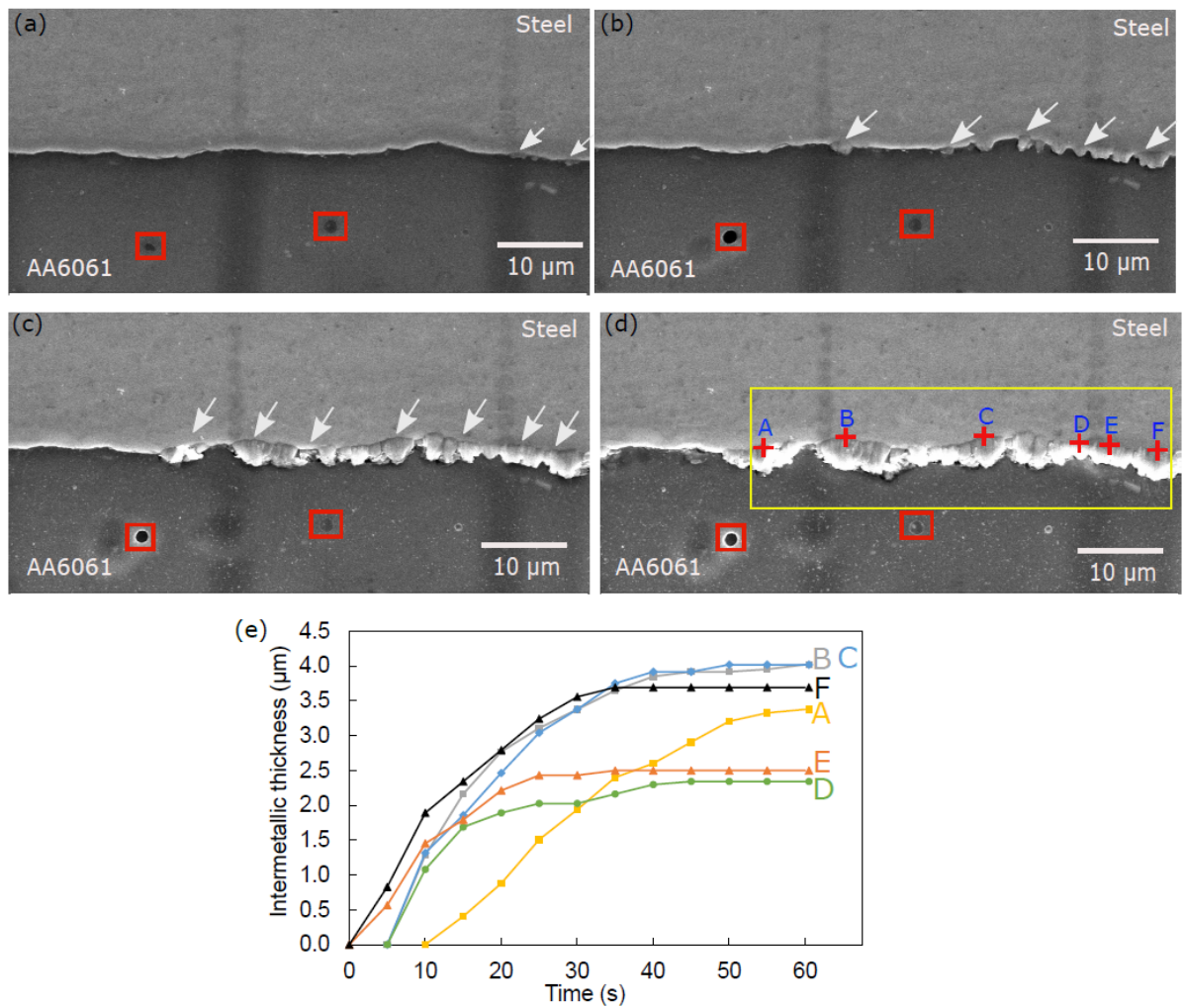

Fig. 2. Sequence of IM nucleation and growth obtained from *in situ* SEM observations at (a) t= 5s, where the onset of nucleation of IM is indicated by white arrows at the interface. The micropores marked by the red squares were used to track the position for further characterization (AFM and EBSD). (b) t = 10s; (c) t= 30s; (d) t= 60s, and (e) time dependent evolution of average IM thickness obtained using image analysis. Due to the low contrast difference between the different IMs formed during this observation and owing to the additional topological effect due to the 3D morphology of the newly formed IMs, the total IM thickness growth is plotted at various locations (marked by A-F in (d)) in (e). White arrows in Fig. 2a-c indicate the nucleation and growth of IMs. The yellow rectangle marked in Fig. 2d refers to the zone corresponding to the EBSD analysis, discussed below in the manuscript. The complete recording of the IM nucleation and growth is provided as Supplementary Material.

After the *in situ* heating experiment, the 3D morphology of the nucleated IMs was characterized by *ex situ* AFM (Fig. 3) when the exact location was identified using red boxed micropores in Fig. 2. The AFM observations corresponding to locations A and B (Figs. 2d and e) show different growth behaviors. 3D views of the AFM maps are provided in Fig. S2 in the Supplementary Material. The AFM analysis revealed that the formation and growth of IMs were accompanied by local expansion/shrinkage. The IM compounds clearly pop up above the original surface of both Al and steel by up to ~1.05 µm due to the local expansion of the sample following solid state reaction [25], while some surrounding zones are clearly below the original surface (up to -1 µm). This effect is more pronounced at location A, corresponding to an IM nucleation site (Fig. 3a), than at location B, corresponding to the significantly grown IM (Fig. 3b). It is clearly seen that at an early stage, the IM starts to pop up out of plane in a random direction (Fig. 3a). This may be attributed to the formation of the θ-phase (as corroborated by further *ex situ* EBSD analysis), which nucleates first by reaction between Al and steel. The developed IM in Fig. 3b clearly shows a preferential growth perpendicular to the interface, which may be attributed to the diffusion-controlled growth of η-phase [26].

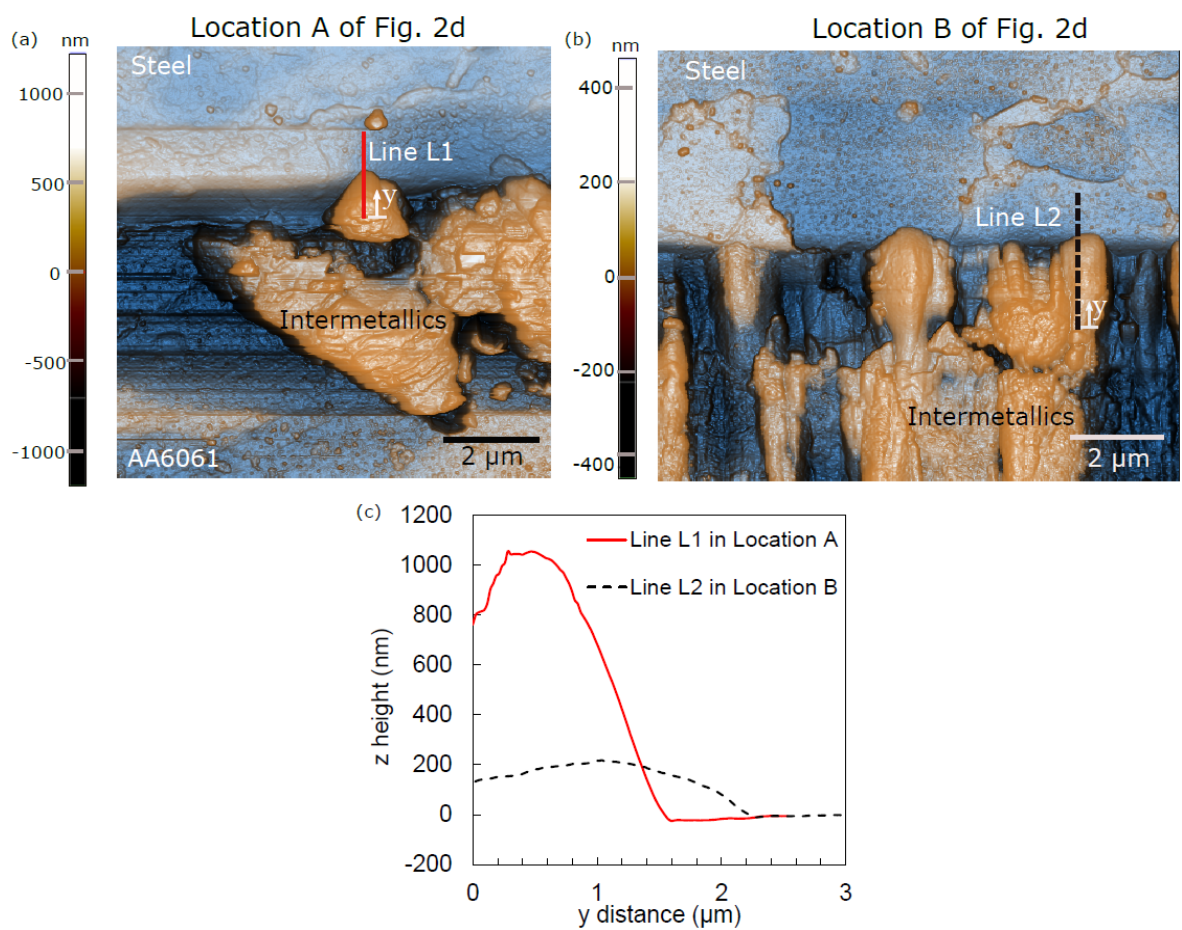

Fig. 3. AFM topography images showing the 3D morphology of the IMs at the Al/Fe interface: (a) IM nucleation (at location A of Fig. 2d) and (b) further directional IM growth across the interface (at location B of Fig. 2d) [color legends indicate the z- height]. (c) Surface profiles corresponding to line L1 (marked in Fig. 3a) and line L2 (marked in Fig. 3b) across the interface. Additional 3D views of the AFM maps are provided in Fig. S2 in the Supplementary Material.

For accurate identification of the formed IM compounds, the same selected interface zones (delimited by a yellow rectangle in Fig. 2d, including locations A-F) were investigated by EBSD phase map, see Fig. 4a. IPF maps of the interface in the x, y, z directions are provided in Fig. S3 in the Supplementary Material. It is worth noting that some variations in morphology are expected due to sample surface preparation. Quantitative analysis of the phase map (Fig. 4a) shows that 79 % of the IM compounds formed at the interface during heat treatment at 596 °C for 60 s correspond to the θ-phase and the remaining 21 % to the η-phase. This confirms that the θ-phase is the 'major phase' in this *in situ* experiment. Moreover, some locations across the interface (marked by black arrows in Fig. 4a) appear fully covered only by the θ-phase, confirming that the θ-phase nucleates first. In some regions, the η-phase also appears at the interface (aqua blue zone in Fig. 4a, e.g. zone B, C and E), confirming that its formation occurs at a later stage. This scenario is also in good agreement with the outcomes of the AFM investigation, where the zone corresponding to location A (Fig. 3a) was assumed to contain only the θ-phase, while location B (Fig. 3b) contains both θ- and η- phases. These observations clearly demonstrate that the terms 'major phase' and/or 'minor phase' are not suitable for the early stages of IMs nucleation and growth. It depends on the stage of growth that the IM has reached and might even be location dependent.

Histograms of grain size distribution for both θ- and η-phases are provided as supplementary in Fig. S4. Both phases are characterized by ultra-fine structure. The earlier formed θ- grains are slightly larger with an average size of 0.22±0.09 μm compared to the η-grains of 0.20±0.09 μm. Both θ- and η-grains are slightly elongated along the phase growth direction, presenting very similar average aspect ratios of 1.84 and 1.86, respectively. It should be noted that these values are much lower compared to those measured on columnar grains of both phases in the well-grown IM layer with the thickness of 38-40 μm [26, 27]. As shown in Fig. 4b the vast majority of η-grains have a strong <001> texture (maximum intensity of 13.3) along the IM growth direction (i.e. perpendicular to the interface), as their growth is controlled by interdiffusion of atoms in that direction. In comparison, the θ-phase controlled by the reaction rate has a weak random texture with a maximum intensity of 3.6.

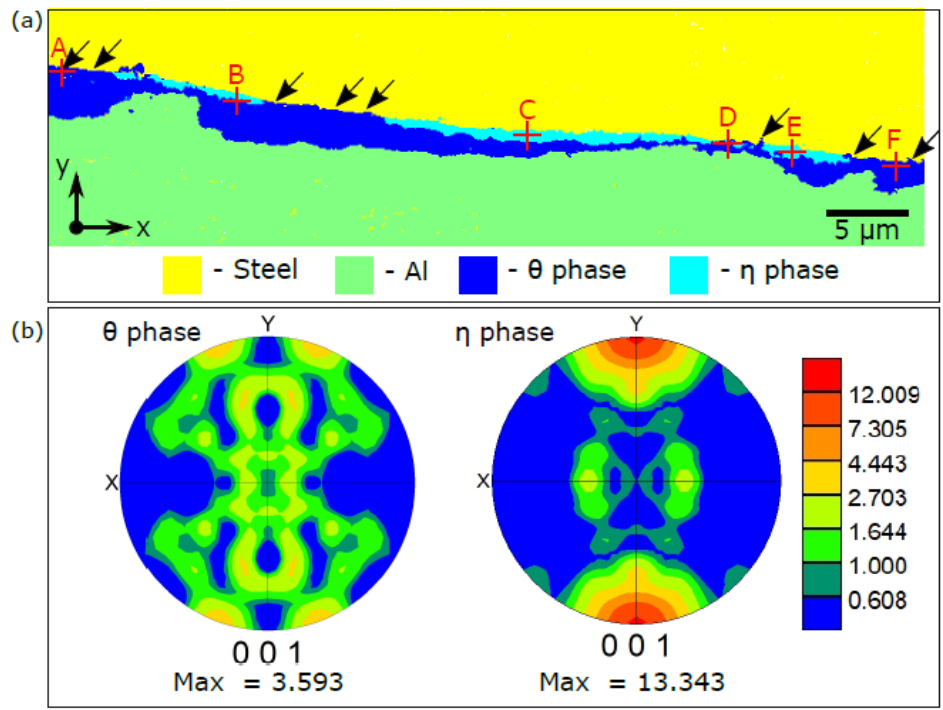

Fig. 4. (a) Phase map of the nucleated IM phases (θ and η) AA6061 and bcc-Fe from EBSD analysis after *in situ* heat treatment corresponding to the area enclosed by the yellow rectangle in Fig. 2d. (b) The <001> texture for both IM phases. [IPF maps of the interface and grain size distributions are given in the Supplementary Material].

To sum up, this is the first-time that a dedicated high temperature *in situ* heating setup is used to successfully capture the onset of nucleation and growth of IMs at the Fe-Al interface. The *in situ* observations enabled to directly observe the location of initiation and to measure the growth rate of IMs. The sample could then be tracked from one technique to the next by location tracking. AFM confirmed that the nucleation and growth of IMs are associated with volume expansion as they are protruding out of plane. EBSD analysis revealed that, in some interface regions, only the θ-phase is observed without emergence of the η- phase. Coupling with the zone tracking identifying nucleation locations of the *in situ* SEM sample, it confirmed that the θ-phase nucleates first. It is concluded that the terms minor and major – phases are not relevant for θ- and η- phases at the early stages of their formation and growth. Indeed, the presence of different IMs depend on the growth time at high temperature, that can vastly change the observations.

**Acknowledgements**

TS acknowledges F.R.S.-FNRS, (Belgium) for his postdoctoral fellowship at UCLouvain and for the research stay at IMDEA Materials Institute (Getafe, Spain). PX acknowledges gratefully the financial support from China Scholarship Council (No. 201606890031, Beijing, China). AS acknowledges the financial support of the European Research Council for a starting grant under grant agreement 716678, ALUFIX project. JMM-A acknowledges MAT4.0-CM project funded by Madrid region under program S2018/NMT-4381.